\documentclass[longauth]{aa}

\usepackage{graphicx}
\usepackage{color}
\usepackage{url}

\usepackage{txfonts}

\begin{document}

   \title{Rapid and multi-band variability of the TeV-bright active nucleus of the galaxy IC 310}

   \author{
  J.~Aleksi\'c\inst{1} \and
 L.~A.~Antonelli\inst{2} \and
 P.~Antoranz\inst{3} \and
 A.~Babic\inst{4} \and
 U.~Barres de Almeida\inst{5} \and
 J.~A.~Barrio\inst{6} \and
 J.~Becerra Gonz\'alez\inst{7} \and
 W.~Bednarek\inst{8} \and
 K.~Berger\inst{7,}\inst{9} \and
 E.~Bernardini\inst{10}
% ...
 %\\
(\textit{The MAGIC Collaboration}),\\
T.~Dauser\inst{34}, P.~Fortin\inst{35}, M.~Kadler\inst{12}, F.~Krau\ss\inst{34,}\inst{12}, 
S.~Wilbert\inst{12}, and J.~Wilms\inst{34}
 }

\date{Received .../ Accepted ...}

\offprints{Dorit Eisenacher (Dorit.Eisenacher@astro.uni-wuerzburg.de) and Pierre Colin (colin@mppmu.mpg.de)}

 \abstract
  {The radio galaxy IC\,310 has recently been identified as a $\gamma$-ray emitter based on observations
at GeV energies with
 \textit{Fermi}-LAT and at very high energies (VHE, E$>$100\,GeV) with the MAGIC telescopes. 
  Originally classified as a head-tail radio galaxy, the nature of this object is subject of controversy since its nucleus shows blazar-like behavior.
  }
  {In order to understand the nature of IC\,310 and the origin of the VHE emission we studied the spectral and flux variability of IC\,310 from the X-ray band to the VHE 
  $\gamma$-ray regime.
  }
 {The light curve of IC\,310 above 300\,GeV has been measured with the MAGIC telescopes from October 2009 to  February 2010.
 Contemporaneous \textit{Fermi}-LAT data (2008--2011) in the 10--500\,GeV energy range were also analyzed.
In the X-ray regime, archival observations from 2003 to 2007 with \textit{XMM-Newton}, \textit{Chandra}, and \textit{Swift}-XRT in the 0.5--10\,keV band were studied.
  }
 {The VHE light curve reveals several high-amplitude and short-duration flares.
  Day-to-day flux variability is clearly present ($>$\,5\,$\sigma$).
  The photon index between 120\,GeV and 8\,TeV remains at the value $\Gamma\sim2.0$ 
  during both low and high flux states. 
  The VHE spectral shape does not show significant variability, whereas the flux at 1\,TeV changes by 
  a factor of $\sim7$.
  \textit{Fermi}-LAT detected only eight $\gamma$-ray events in the energy range 10\,GeV--500\,GeV in three years of observation. The measured photon index of 
  $\Gamma=1.3\pm0.5$ in the \textit{Fermi}-LAT range is very hard.
  The X-ray measurements show strong variability in both flux and photon index. The latter varied from $1.76\pm0.07$ to $2.55\pm0.07$.
  }
 {The rapid variability measured in $\gamma$-rays and X-rays confirms the blazar-like behavior of IC\,310. The multi-TeV
$\gamma$-ray emission seems to originate from scales of less than 80 Schwarzschild radii (for a black hole mass of $2\times\,10^{8}$\,M$_{\odot}$) within the compact core of its FR~I radio jet 
with orientation angle $10^{\circ} - 38^{\circ}$.
The spectral energy distribution resembles that of an extreme blazar, albeit the luminosity is more than two orders of magnitude lower. 
}


   \maketitle
%
%________________________________________________________________

\section{Introduction}

The galaxy IC\,310 (redshift of $z=0.0189$; \cite{bernardi02}) is one of the brightest objects of the Perseus cluster of galaxies at radio frequencies as well as at X-ray energies.
It has been detected in the high energy $\gamma$-ray band above 30\,GeV with the \emph{Fermi} Large Area Telescope (\emph{Fermi}-LAT; \cite{neronov}) 
and above 260\,GeV with the MAGIC telescopes (Aleksi\'c et al. \cite{aleksic10}).

IC\,310 was originally classified as a head-tail radio galaxy (HTRG; \cite{sijbring}) based on the radial alignment of its radio jet on kiloparsec scale with the radially oriented pressure 
gradient in the surrounding intracluster medium (ICM). 
The radio morphology of such radio galaxies consists of a bright ``head'', located at the core of the host galaxy, and ``tails'' pointing away from the 
center of the cluster. In the case of IC\,310, the jet bending was assumed to be large, giving rise to the classification as a narrow-angle 
HTRG (\cite{ryle}; \cite{sijbring}; \cite{miley80}; \cite{lal}; \cite{feretti}).

A recent detailed investigation of the radio structure of IC\,310 using VLBA observations (Kadler et al. \cite{kadler}) 
questions this classification. Inside the ``head'', a parsec-scale core-jet structure was detected with no significant detection of a counter jet.
The parsec-scale jet appears oriented in the same direction as the kiloparsec structure (the ``tail'').
This morphology disagrees with the classification as a HTRG since there is no indication for a jet bending process that determines the direction of the tail.
Instead, the missing counter-jet implies that Doppler-boosting or other anisotropies play a significant role in this object. 
Relativistic beaming is commonly found in blazars,
i.e., active galactic nuclei (AGN) with their jets pointing very close to the line of sight.
Thus, IC\,310 appears to be more similar to other AGN detected in the very high energy regime (VHE, $>$100\,GeV) that are mostly blazars (\cite{ackermann11}).

On the other hand, if there is no bend, the $\sim$400\,kpc scale (projected) radio jet (\cite{sijbring}) would indicate at least a moderately large angle between the jet and the line of sight. 
IC\,310 would then belong to the same class as the three other ``non-blazar'' AGN detected at VHE, 
M\,87 (\cite{aharonian03}; \cite{aharonian06}; \cite{acciari08}; \cite{albert08}), Centaurus\,A (\cite{aharonian09}), and NGC\,1275 (\cite{aleksic12a}) 
which are Fanaroff-Riley~I (FR~I) radio galaxies.
Hence, IC\,310 may be considered either the closest TeV blazar or the brightest radio galaxy at TeV energies.

The classification of IC\,310 in a transitional population between BL Lac objects and FR~I radio galaxies was suggested by \cite{rector99}, based on its optical,
radio and X-ray properties. Its weak optical emission lines are similar to those typically found in FR~I radio galaxies but the non-thermal continuum from radio to 
the X-ray range is comparable to a low-luminosity BL Lac (\cite{owen96}) .
In X-rays, the object is strongly dominated by the non-thermal point-like emission coincident with the radio ``head'' (\cite{schwarz92}; \cite{rhee94}; \cite{sato}).
A faint X-ray halo has been observed by \textit{Chandra} extending in the direction of the radio tail (\cite{dunn}).

Aleksi\'c et al. (\cite{aleksic10}) reported the detection of IC\,310 during observations of the Perseus cluster with MAGIC taken place between 
2008 November and 2010 February. A hard spectrum ($F\propto E^{-\Gamma}$ with $\Gamma=2.0\pm0.14$) between 150\,GeV and 7\,TeV and indications 
for flux variability on time scale of months and years were reported. 
The variability time scales of years is confimed by the non-detection of the source reported in 2010 August - 2011 February (\cite{aleksic12a})
 as well as by a re-detection in more recent MAGIC observations (ATel \#4583).

We present a re-analysis of the MAGIC stereo-observation taken during the period of strong activity between 2009 October and 2010 February
allowing faster sampling of the light curve and more accurate spectral analysis. We also investigate $\gamma$-ray data of the \emph{Fermi}-LAT instrument between 2008--2011
and archival X-ray observations with \textit{XMM-Newton} (2003), \textit{Chandra} (2004 and 2005) and \textit{Swift}-XRT (2007).
In Sect. 2, the $\gamma$-ray and X-ray observations and methods of data extraction and analysis are presented, while the resulting light curves
and energy spectra are reported in Sect. 3.  Finally, the observational findings are discussed in Sect.~4.

We adopt a cosmology with $\Omega_{\mathrm{m}}=0.27$, $\Omega_{\Lambda}=0.73$ and $H_{0}=71$\,km\,s$^{-1}$\,Mpc$^{-1}$.
%__________________________________________________________________

\section{Data: observations and analysis}

\subsection{VHE $\gamma$-ray data: The MAGIC telescopes}

VHE observations of IC\,310 were carried out with the MAGIC telescopes which are two Imaging Atmospheric Cherenkov Telescopes
located on the island of La Palma at an altitude of 2200\,m.
Both telescopes consist of a mirror dish of 17\,m diameter associated with a fast imaging camera of $3.5^{\circ}$ field of view.
The trigger threshold is $\sim$50\,GeV and the sensitivity above 290\,GeV (in 50\,h) is $\sim$0.8\% of the Crab Nebula flux
with an angular resolution better than $0.07^{\circ}$ (Aleksi\'c et al. \cite{aleksic12b}).

From 2009 October to 2010 February, the Perseus cluster was observed with MAGIC in the so-called wobble mode, i.e. pointing alternatively to
two positions $0.4^{\circ}$ away from the center of the cluster (NGC\,1275). Since IC\,310 is located $0.6^{\circ}$ from the cluster center,
it appeared as an off-axis source in these observations. The position of the source is $0.25^{\circ}$ and $1^{\circ}$ away
from the camera center in the two wobble positions, respectively. In our previous publication on IC\,310 (Aleksi\'c et al. \cite{aleksic10}) only data of the closest wobble position were used
to construct the light curve and the spectrum. Here, we used an improved analysis method that can handle the data of 
both wobble positions and hence provides more accurate results.

After data quality selection mainly based on the atmospheric conditions, the data sample corresponds to 43.3\,h effective time, $t_{\mathrm{eff}}$.
The effective time of single MAGIC observations is reported in Tab.~\ref{table:1}.
The calibration, image cleaning, parameterization, and event reconstruction as well as the gamma/hadron separation were performed with the standard analysis 
software MARS described in \cite{moralejo09}. The background estimation is done separately for each wobble position. For the closer wobble position 2 OFF (signal-free) regions 
at $0.25^{\circ}$ away from the camera center were used and for the wobble position further away 5 OFF regions were chosen with an offset of $1^{\circ}$.
The $\gamma$-ray signal is calculated by subtracting the estimated background from the on-source region events.
The effective area, $A_{\mathrm{eff}}$, of MAGIC strongly depends on the distance from the camera center. It is estimated separately for each wobble position
using Monte Carlo simulations of $\gamma$ rays initiated at the same distance from the camera center as the source position.

The spectrum and the light curve from both wobble positions are combined by calculating the sum, over two wobble positions, of the total number of excess events, 
$N_{\mathrm{ex}}$, measured during each wobble observation weighted by the corresponding effective area and effective time.

Because IC\,310 is not located in the central position, the systematic errors can be higher than reported for the
standard $0.4^{\circ}$ wobble observation by Aleksi\'c et al. (\cite{aleksic12b}).
In order to study the systematic effects, we analyzed Crab Nebula observations taken at different offsets from the camera center (from $0.2^{\circ}$ 
to $1.4^{\circ}$) with the same analysis chain. The Crab spectra measured at different offsets, e.g. with $0.2^{\circ}$ and $1^{\circ}$, are in good agreement with the standard observation result.
The systematic errors on the flux normalization and photon index are estimated to be below $\sim$17\% and $\sim$0.2, respectively 
(instead of 11\% and 0.15 for standard wobble observations; Aleksi\'c et al. \cite{aleksic12b}). The systematic uncertainty on the energy scale is evaluated to be 15\% 
(Aleksi\'c et al. \cite{aleksic12b}).

\subsection{HE $\gamma$-ray data: \textit{Fermi}-LAT}

The {\it Fermi}-LAT is a pair-conversion telescope sensitive to photons between 20~MeV and several hundred GeV (\cite{atwood}; \cite{ackermann12}). 
Since August 5, 2008 it has operated primarily in sky survey mode, scanning the entire sky every three hours. The data used in this paper were taken between 
August 5, 2008 and July 31, 2011 (MJD 54683--55773), overlapping the MAGIC observations.  
The \textit{Fermi}-LAT results presented here were obtained with the analysis pipeline used to produce the ``The First Fermi-LAT Catalog of Sources Above 10\,GeV'', designated 1FHL (\cite{Ackermann2013}).
This analysis was performed with the ScienceTools software package version v9r26p02. The results presented here were obtained with ``Clean" class events in the energy 
range 10--500~GeV from the region centered at (R.A., Dec.) = ($52.921^{\circ}$, $41.634^{\circ}$) (J2000), and has a radius of $6.198^{\circ}$. Only data for time periods when 
the spacecraft rocking angle was less than 52$^{\circ}$ were used, and events with zenith angles larger than 105$^{\circ}$ were excluded in order to reduce the contamination 
from Earth limb $\gamma$-rays, which are produced by cosmic rays interacting with the upper atmosphere.

In the analysis, the Galactic and extragalactic diffuse backgrounds were parameterized with the files {\tt gal\_2yearp7v6\_v0.fits} and {\tt iso\_p7v6clean.txt}, 
which are publicly available\footnote{See \url{http://fermi.gsfc.nasa.gov/ssc/data/access/lat/BackgroundModels.html}}.  Because of the relatively small size of the region, 
the limited photon count, and the relatively small effective energy range (most photons cluster in the energy range 10--100~GeV), there is some degeneracy in the simultaneous 
characterization of the Galactic and isotropic diffuse components, so we fixed the normalization of the isotropic component to the best-fit value over the entire sky, and left free the normalization of the Galactic component.  

The region analyzed contains only two sources detected above $>$10~GeV, which based on close positional agreement we have associated with IC~310 and NGC~1275. 
The source positions used in the spectral fit were the optimized positions from the LAT analysis, which are ($49.169^{\circ}$, $41.322^{\circ}$) 
for IC~310 and ($49.977^{\circ}$, $41.501^{\circ}$) for NGC~1275. These positions differ from the actual source positions by $0.01^{\circ}$ and $0.02^{\circ}$ respectively for 
IC~310 and NGC~1275, which are well within the 95\% confidence regions (error ellipses) for these two objects: $0.09^{\circ}$ and $0.03^{\circ}$. The source associated 
with NGC~1275 is $\sim$20 times brighter than IC~310 at GeV energies and separated by only $\sim$\,$0.65^{\circ}$.  We note that above 10~GeV the 68\% containment radius 
of the point-spread function of the LAT is about $0.2^{\circ}$, which means that the sources are resolved, and in particular the spectrum of IC~310 can be separately measured.

We used simple power-law models to characterize the spectra of the sources. The spectral fitting was performed with the binned likelihood method using the P7\_V6\_CLEAN instrument response functions 
(see Ackermann et al. \cite{ackermann09}), ten bins per decade in energy starting at 10~GeV, and an angular binning of $0.05^{\circ}$ and $0.1^{\circ}$ for $\gamma$-rays that converted in the thin and thick tungsten 
layers of the tracker, respectively.  Following the global fitting over the full energy range, we extracted photon fluxes in three energy bands: 10--30 GeV, 30--100 GeV and 100--500 GeV by fixing the photon 
indices (to those from the overall spectral fit) and leaving free only the normalizations in the fits. Because of the low statistics, the flux errors are strongly dominated by Poisson fluctuations and so are not symmetric. 
In this manuscript we report separate 1 $\sigma$ uncertainties toward low/high fluxes obtained via MINOS in the MINUIT package.

\subsection{X-ray data: Chandra, XMM-Newton, Swift-XRT}

Two \textit{Chandra} observations of IC\,310 were taken with the Advanced CCD Imaging Spectrometer (ACIS) in the 0.5--8\,keV band (ACIS-I Observation ID 5596 and 5597).
The data sets are not affected by pile-up or detector heating.
The observations were made within four months of each other: 
December 26, 2004 with effectively 25.2\,ks for Obs.\,ID 5597 and March 23, 2005 with 1.5\,ks for Obs.\,ID 5596.
The data were analyzed with the Chandra Interactive Analysis of Observations (CIAO 4.4) software using Version 4.4.8 of the calibration files.
The extraction radii were chosen to be 4.92$''$ for Obs.\,ID 5597 and 2.46$''$ for Obs.\,ID 5596.

One observation of IC\,310 with an exposure of 22.6\,ks was taken on February 26, 2003 (Obs.\,ID 0151560101) with \textit{XMM-Newton}. 
Here, we analyzed only data taken with the pn detector of the European Photon Imaging Camera (EPIC-pn) covering the energy range 0.2--15\,keV (\cite{strueder01}).
Data were reduced using the \textsl{XMM-Newton} Software Analysis System (SAS v.11.0.0) 
and the newest calibration files. For the source spectra, we used a circular region of 30$''$ radius, centered on the source. 
The background was extracted from an equally large circle, but located outside any distinct source of radiation. Only single and double 
events were used to derive the spectrum.

Swift consists of several instruments covering a broad spectral range 
including the X-ray telescope (XRT; \cite{burrows}) operating in the 0.2--10\,keV band.
The observation of IC\,310 was performed on February 19, 2007 with an effective observation time of 4.1\,ksec.
The data were reduced with XRTPIPELINE (V 0.12.6).
For the source region, a circle with a radius of 47$''$ was used. The background region was created using an annulus centered on the
source coordinates with an inner radius of 118$''$ and an outer radius of 165$''$.
The extraction of events was done in XSELECT (V2.4b), using event grades 0--12.  

The X-ray data analysis was performed with the \textsl{Interactive Spectral Interpretation System} (\cite{houck00}). 
In order to allow for a reasonable spectral fitting, we re-binned the data according to a signal-to-noise criteria for each bin (eight for \textsl{Swift} 
and \textsl{XMM-Newton}, and four in the case of the two \textsl{Chandra} observations). For the spectral fitting, we used data between 0.5\,keV and 10\,keV for 
all three instruments. The photon index $\Gamma$ and the column density of the neutral hydrogen absorption was determined by the spectral fitting. As IC\,310 is located $0.6^{\circ}$
away from the center of the cluster, the contribution of the thermal ICM emission to the X-ray signal is very small, and therefore, it was not necessary to include it when 
modeling the data.

%______________________________________________________________
\section{Results}
\subsection{VHE light curve}
%___________________________________________________________
 
\begin{figure*}
   \centering
      \includegraphics[width=16cm]{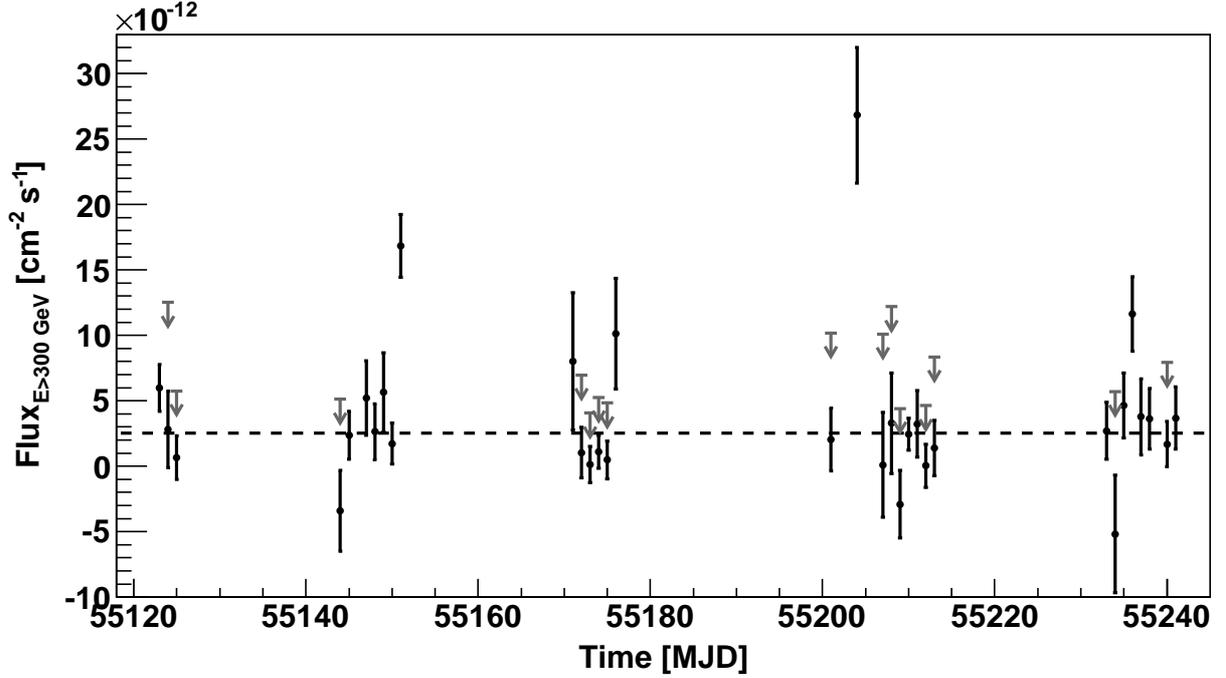}
     \caption{Light curve of IC\,310 above 300\,GeV from 2009 October to 2010 February.
   The arrows are 95\% confidence level upper limits calculated for days compatible with no signal.
   The dashed line shows the constant flux fit to all flux points not taking into account the upper limits ($\chi^2$/d.o.f = 102/32).}
              \label{LC}%
    \end{figure*}
%______________________________________________________________

The integral VHE $\gamma$-ray flux of IC\,310 has been derived assuming a
differential spectrum for the calculation of the effective area following a power-law with a photon index
$\Gamma=-2.0$ as found in Aleksi\'c et al. (\cite{aleksic10}). However, the dependence of the resulting light curve on the index is only minor. The flux above 300\,GeV
measured during 33 individual days between 2009 October and 2010 February is shown Fig.~\ref{LC} and 
listed in Table~\ref{table:1}.
Integral upper limits have also been calculated for days showing an excess below one standard deviation ($\sigma$). The upper limits are determined by applying model 4 of \cite{rolke}, 
using a confidence level (c.l.) of 95\% and 30\% systematic uncertainty.
The mean flux above 300\,GeV during this period is $\Phi_{\mathrm{mean}}=(3.62 \pm 0.40)\times10^{-12}$\,cm$^{-2}$s$^{-1}$ which is in good agreement with the mean flux
reported previously, $\Phi_{\mathrm{mean}}=(3.1 \pm 0.5)\times10^{-12}$\,cm$^{-2}$s$^{-1}$, using only the $0.25^{\circ}$ wobble position (Aleksi\'c et al. \cite{aleksic10}).
The best fit of the flux points, i.e. not including the upper limits, with a constant flux (dashed line, $\Phi_{\mathrm{CstFit}}=(2.52 \pm 0.37)\times 10^{-12}$\,cm$^{-2}$s$^{-1}$) has a $\chi^2$ test value of $102$ for 
32 degrees of freedom corresponding to a probability of $3.10\times\,10^{-9}$ that the source is not variable. Note, that since the high $\chi^2$ value 
rejects the validity of the constant flux fit, this value cannot be interpreted as the average flux of the source in the observed time span. The variability of the source is thus detected with 
a confidence level of 5.8\,$\sigma$.

\begin{table}
\caption{Results from individual days of MAGIC observations.}             
\label{table:1}     
\centering                        
\begin{tabular}{c c c c}       
\hline\hline
   used data\tablefootmark{a}             &MJD start     & $t_{\mathrm{eff}}$ & $F_{\mathrm{E}>300\,\mathrm{GeV}}$\tablefootmark{b} \\    
  & & [h] & [$10^{-12}$\,ph\,cm$^{-2}$\,s$^{-1}$] \\
\hline 
   all data              &               	         & 43.32         & $3.62\pm0.40$                  \\
\hline 
  2009-10-19      & 55123.02        		 & 2.43          & $5.97\pm1.79$              \\
  2009-10-20     & 55124.05        		 & 1.11          & $(2.80\pm2.93)<12.52$   \\
  2009-10-21     & 55125.02         		 & 1.58          & $(0.67\pm1.67)<5.74$  \\
  2009-11-09     & 55143.96         		 & 0.80          & $(-3.42\pm3.09)<5.11$  \\
  2009-11-10     & 55144.97         		 & 1.19          & $2.36\pm1.83$              \\
  2009-11-12     & 55146.98         		 & 0.99          & $5.21\pm2.86$              \\
  2009-11-13     & 55147.98         		 & 1.17          & $2.64\pm2.14$              \\
  2009-11-14     & 55148.98         		 & 1.11          & $5.65\pm3.03$              \\
  2009-11-15     & 55149.94         		 & 2.61          & $1.73\pm1.57$              \\
  2009-11-16     & 55150.94         		 & 2.34          & $16.83\pm2.40$    \\
  2009-12-06     & 55170.89         		 & 0.44          & $8.01\pm5.25$              \\
  2009-12-07     & 55171.89         		 & 1.44          & $(1.03\pm1.94)<6.95$   \\
  2009-12-08     & 55172.88         		 & 2.51          & $(0.13\pm1.39)<4.06$   \\
  2009-12-09     & 55173.89         		 & 3.18          & $(1.10\pm1.25)<5.25$   \\
  2009-12-10     & 55174.88         		 & 2.40          & $(0.49\pm1.44)<4.83$   \\
  2009-12-11     & 55175.88         		 & 0.54          & $10.13\pm4.23$             \\
  2010-01-05     & 55200.83         		 & 0.81          & $(2.05\pm2.40)<10.02$  \\
  2010-01-08     & 55203.89         		 & 0.78          & $26.84\pm5.19$    \\
  2010-01-11     & 55206.87         		 & 0.62          & $(0.09\pm4.00)<10.09$  \\
  2010-01-12     & 55207.90         		 & 0.61          & $(3.29\pm3.85)<12.19$   \\
  2010-01-13     & 55208.88         		 & 0.86          & $(-2.91\pm2.59)<4.42$  \\
  2010-01-14     & 55209.84         		 & 1.47          & $2.44\pm1.22$              \\                                               
  2010-01-15     & 55210.83         		 & 1.46          & $3.23\pm2.54$              \\
  2010-01-16     & 55211.83         		 & 1.90          & $(0.03\pm1.66)<4.63$   \\
  2010-01-17     & 55212.84         		 & 1.22          & $(1.40\pm2.11)<8.33$   \\ 
  2010-02-06     & 55232.84        		 & 1.22          & $2.71\pm2.18$             \\
  2010-02-07     & 55233.87         		 & 0.60          & $(-5.19\pm4.50)<5.71$  \\
  2010-02-08     & 55234.85         		 & 0.90          & $4.65\pm2.47$              \\
  2010-02-09     & 55235.85         		 & 1.38          & $11.65\pm2.84$    \\
  2010-02-10     & 55236.87        		 & 0.62          & $3.77\pm2.91$              \\
  2010-02-11     & 55237.86         		 & 0.87          & $3.62\pm2.32$              \\
  2010-02-13     & 55239.84         		 & 0.92          & $(1.69\pm1.74)<7.95$   \\
  2010-02-14     & 55240.84         		 & 1.24          & $3.68\pm2.38$              \\

\hline                                  
\end{tabular} 
\tablefoot{
\tablefoottext{a}{Dates in MAGIC night notation.}
\tablefoottext{b}{Measured flux above 300\,GeV in units of $10^{-12}$\,ph\,cm$^{-2}$s$^{-1}$. Upper limits are given with 95\% confidence level.}
}
\end{table}

Three days show a flux $>$\,3$\sigma$ above the constant flux fit: November 16, 2009 (MJD = 55151), January 8 (MJD = 55204) and February 9, 2010 
(MJD = 55236)\footnote{Observations were taken around midnight UTC. The dates given in this paper correspond to the following day. 
Exact observation times are given in Table~\ref{table:1}.}.
The mean flux during these three days is $(1.60 \pm 0.17)\times10^{-11}$\,cm$^{-2}$s$^{-1}$, which is more than six times higher than the constant fit flux.
Excluding these three days, the light curve is compatible with a constant flux (probability of 36\%).
%___________________________________________________________
   \begin{figure}
   \centering
   \includegraphics[width=8cm]{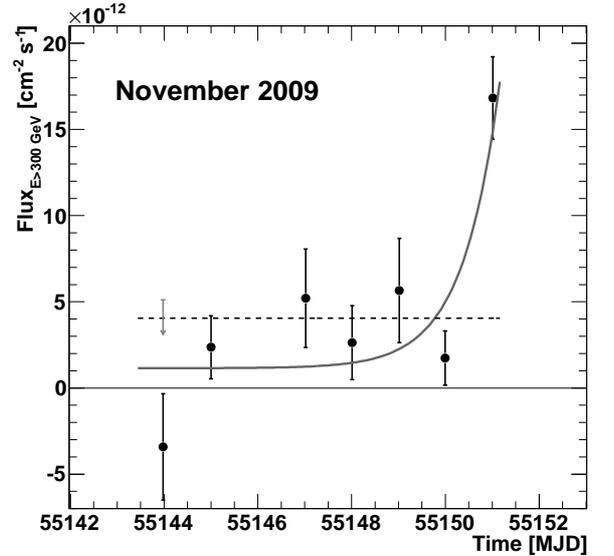}
   \caption{Daily light curve of IC\,310 above 300\,GeV in 2009 November.
   The black dashed line shows the best constant fit of the data.
   The thick grey line shows the constant plus an exponential component
   corresponding to the doubling time of $\tau_{\mathrm{Nov, UL}}$=0.55\,d providing an acceptable fit of the data (5\% of probability).
}
   \label{LC-zoom}	
   \end{figure}
%______________________________________________________________
The three flares are isolated and seem to be restricted to a single day bin.
The most significant one happened in 2009 November. Figure~\ref{LC-zoom} shows a zoom-in of the daily light curve for this month.
IC\,310 was observed every day from 9 to 16 except for November 11.
The probability of a constant flux during this period is 10$^{-6}$. 
The flare appeared during the last day of observation in November and no evidence of an increased flux was seen before. 
The variability time scale might be shorter than the daily-scale sampling of the light curve but, due to limited statistics intra-night variability could not be established.

The characteristic flux doubling time during a flare can be estimated by fitting the data with a constant plus an exponential increase.
However, in the night just before the flare in November the flux was particularly low and the best fit doubling time goes to zero.
To estimate the flux-doubling time, we fit the data with fixed doubling times and calculate the probability of 
each hypothesis with the $\chi^2$ method. The largest flux-doubling time providing a fit probability 
above 5\% is $\tau_{\mathrm{Nov, UL}}$=0.55\,d (thick line in Fig.~\ref{LC-zoom}).

The 2010 January and February flares provide much less constraining results because of a sparse time coverage in January and a much less significant flare in February. 
For the rest of the paper, we consider only the upper limit obtained from the November data.

\subsection{VHE spectra}

\begin{figure}
   \centering
      \includegraphics[width=9cm]{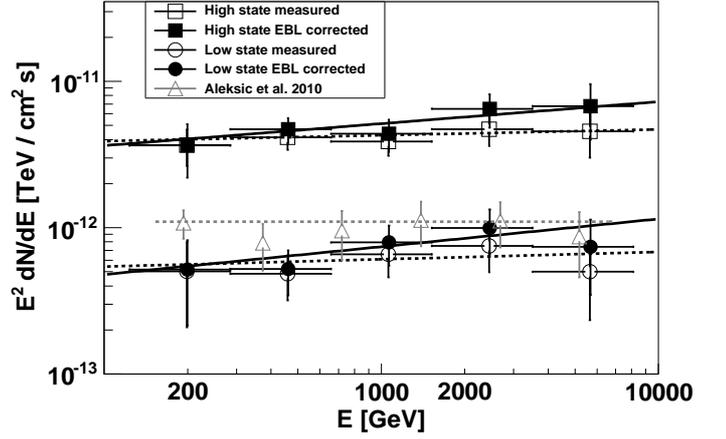}
     \caption{Measured (open markers) and EBL-absorption corrected (filled markers) spectral energy distribution for IC\,310 obtained by MAGIC in high and low states 
     together with their power-law fits in dashed line (measured) and solid line (EBL corrected). Fit parameters are given in Table~\ref{table:1}.
     For comparison, we show the result reported in Aleksi\'c et al. \cite{aleksic10} (grey triangles, without EBL de-absorption) for the whole period using only 
    the closest pointing observations.}
              \label{SEDEBL}%
    \end{figure}

For the spectral analysis, we split the MAGIC data set into two samples corresponding to different $\gamma$-ray emission states.
We define a ``high state'' containing the three measurements with $F_{\mathrm{E}>300\,\mathrm{GeV}}>1.1\times10^{-11}$\,cm$^{-2}$\,s$^{-1}$ (MJD 55151, 55204 and 55236, $t_{\mathrm{eff}}$=4.5\,h), which
are the three days with the single-night flux measurements that deviate by more than 2$\sigma$ with respect to the average flux, as shown in Fig.~\ref{LC}. The remaining observations ($t_{\mathrm{eff}}$=38.8\,h)
were grouped to produce a spectrum for the non-flaring or ``low state''. The reconstructed spectra between 120\,GeV and 8.1\,TeV for both states are shown in Fig.~\ref{SEDEBL}.
Due to the absorption of VHE $\gamma$ rays by the Extragalactic Background Light (EBL) through pair creation, the observed spectra of extragalactic 
sources are softened. We corrected the spectra for the EBL absorption according to different EBL models (Dominguez et al. \cite{dominguez11}; \cite{franceschini08}; \cite{kneiskedole10}) 
which all provide comparable results within the systematic errors. The EBL correction applied here uses the model by Dominguez et al. (\cite{dominguez11}).
Due to the proximity of IC\,310 ($z=0.0189$), the effect of the absorption is relatively modest, reducing the flux above 1\,TeV by 15--20\% and steepening the photon 
index by $\sim$0.1.
 
Both observed and de-absorbed spectra can be well described by a simple power law for both emission states:
\begin{equation}
\frac{\mathrm{d}F}{\mathrm{d}E}=f_0\times\left(\frac{E}{1\mathrm{TeV}}\right)^{-\Gamma}\left[\frac{10^{-12}}{\mathrm{cm}^{2}\mathrm{s}\,\mathrm{TeV}}\right].
\end{equation}
Results for the flux normalization at 1\,TeV, $f_0$, and the photon index, $\Gamma$, are summarized in Table~\ref{table:VHESpecta}.
Changes in $f_0$ by a factor of $\sim$7 between the low and high states had been observed, but no significant change in the photon index.

\begin{table}
\caption{Results of power-law fit of the 0.12--8.1\,TeV spectra measured with MAGIC.}            
\label{table:VHESpecta}     
\centering                         
\begin{tabular}{c c c c}       
\hline\hline
state    &&$f_{0}\pm f_{\mathrm{stat}}\pm f_{\mathrm{syst}}$ &$\Gamma\pm\Gamma_{\mathrm{stat}}\pm\Gamma_{\mathrm{syst}}$\\
         &&$\times10^{-12}[\mathrm{TeV}^{-1}\,\mathrm{cm}^{-2}\,\mathrm{s}^{-1}]$&\\
\hline
high     &observed   &$4.28\pm0.21\pm0.73$     &$1.96\pm0.10\pm0.20$\\
	 &intrinsic  &$5.14\pm0.28\pm0.90$     &$1.85\pm0.11\pm0.20$\\
low      &observed   &$0.608\pm0.037\pm0.11$   &$1.95\pm0.12\pm0.20$\\
	 &intrinsic  &$0.741\pm0.045\pm0.14$   &$1.81\pm0.13\pm0.20$\\
\hline      
\end{tabular}

\end{table}

\subsection{Fermi-LAT results}

The three year data taken with $Fermi$-LAT resulted in the detection of IC\,310 above 10\,GeV with a test statistic of $\mathrm{TS}=27.0$ (4.5\,$\sigma$).
The integrated flux $F_{10-500}$ between 10 and 500\,GeV is measured to be $(6.9\pm3.3)\times10^{-11}$\,$\mathrm{cm}^{-2}\,\mathrm{s}^{-1}$.
The spectrum in this energy band can be fitted by the following power-law formula with a very hard photon index of $\Gamma=1.3\pm0.5$:
\begin{equation}
\frac{\mathrm{d}F}{\mathrm{d}E}=\frac{F_{10-500}(-\Gamma+1)E^{-\Gamma}}{E_{\mathrm{max}}^{-\Gamma+1}-E_{\mathrm{min}}^{-\Gamma+1}}\left[\frac{1}{\mathrm{cm}^{2}\mathrm{s}\,\mathrm{GeV}}\right].
\end{equation}
Here, $E_{\mathrm{min}}$ and $E_{\mathrm{max}}$ are the lower and the upper boundary of the energy bins in GeV, respectively.

\begin{figure*}
   \centering
             \includegraphics[width=13cm]{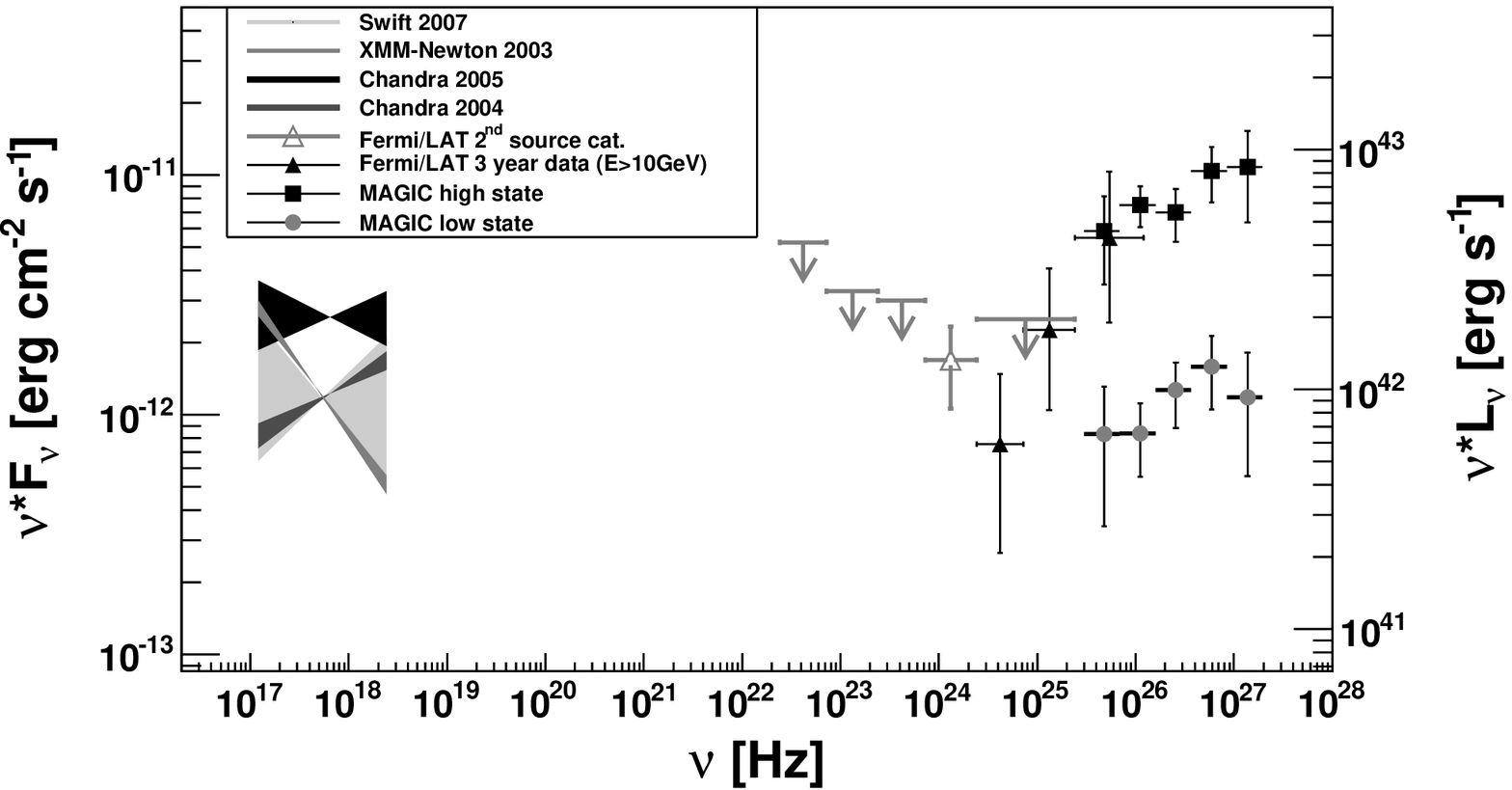}
	     \caption{Spectral energy distribution from X-rays to VHE $\gamma$ rays. 
              In X-ray, butterflies of the observations are shown.
              The gray, open triangle and the gray upper limits are obtained from the \textit{Fermi}-LAT second source catalog (\cite{nolan12}).
              Filled triangles depict the results from the dedicated high-energy analysis reported here.
              The MAGIC results (EBL corrected) for the high and low states are marked as
	      full squares and full circles, respectively. The corresponding apparent luminosity is given at the right axis. For the calculation, the luminosity 
	      distance of 81\,Mpc obtained from the Cosmology Calculator (Wright \cite{wright06}) has been used.}
              \label{FigSED}
    \end{figure*}

The estimated fluxes in three energy bands (10--30\,GeV, 30--100\,GeV and 100--500\,GeV) are shown in Fig.~\ref{FigSED}. Because of the low statistics, the flux errors are strongly
dominated by Poisson fluctuations. Thus, they are asymmetric. In this manuscript we report separate 1\,$\sigma$ uncertainties toward low/high fluxes obtained via MINOS in the MINUIT package.

In $Fermi$-LAT data, seven of the eight photons detected above 10\,GeV arrived within the first 1.5 years of observation (see arrival times in Table~\ref{table:FermiEvents}).
This also suggests variability even if the low statistics do not allow a certain conclusion.
We did not find any relation between the $Fermi$-LAT photon arrival times and the MAGIC flares.

\begin{table}
\caption{Arrival timestamps and energies of $\gamma$-candidates above 10\,GeV (from the first three years of \textit{Fermi}-LAT accumulated data) from a $0.3^{\circ}$ radius circle centered at 
the position of IC\,310. The energy resolution of the \textit{Fermi}-LAT instrument at the energies reported for these events is about 10\% (\cite{ackermann12}). }            
\label{table:FermiEvents}     
\centering                          
\begin{tabular}{c c}      
\hline\hline
MJD     &Energy [GeV]\\
\hline
54720.03	&96.4\\
54833.95	&112.1\\
54846.64	&22.2\\
54972.38	&12.6\\
55081.11	&39.0\\
55118.56	&148.3\\
55247.01	&12.1\\
55462.98	&46.3\\
\hline      
\end{tabular}
\end{table}

\begin{table*}
\caption{Results of the analyzed X-ray observations.}        
\label{table:2}     
\centering                       
\begin{tabular}{c c c c c c c c}       
\hline\hline
instrument    &date         &exposure &$F_{0.5-2\,\mathrm{keV}}$\tablefootmark{a}         &$F_{2-10\,\mathrm{keV}}$\tablefootmark{b}          &$\Gamma$\tablefootmark{c} &$N_{\mathrm{H}}$\tablefootmark{d} &$\chi^2$/d.o.f.\\
              &[MJD]        &[ks]     &[10$^{-3}$\,keV\,s$^{-1}$\,cm$^{-2}$]&[10$^{-3}$\,keV\,s$^{-1}$\,cm$^{-2}$]& 	      		     &[10$^{22}$\,cm$^{-2}$]            &               \\
\hline
\textit{XMM-Newton}    &52697        &22.6     &1.007$\pm$0.012				  &0.828$_{-0.040}^{+0.026}$		      &2.55$_{-0.04}^{+0.07}$ &0.146$_{-0.008}^{+0.016}$ &124/104\\
\textit{Chandra} ObsID 5596  &53456        &1.5      &1.77$\pm$0.13 				  &2.5$\pm$0.4				      &2.01$\pm$0.20          &0.07$_{-0.07}^{+0.08}$    &62/78\\
\textit{Chandra} ObsID 5597  &53363        &25.2     &0.656$\pm$0.019				  &1.39$\pm$0.08			      &1.76$\pm$0.07          &0.089$_{-0.027}^{+0.028}$ &97/78\\
\textit{Swift}-XRT     &54152        &4.1      &0.82$\pm$0.10				  &1.2$_{-0.5}^{+0.6}$			      &2.0$_{-0.4}^{+0.5}$    &0.07$_{-0.07}^{+0.13}$    &12/16\\
\hline      
\end{tabular}
\tablefoot{
\tablefoottext{a}{Measured flux between 0.5 and 2\,keV determined by a simple power-law fit.}
\tablefoottext{b}{Measured flux between 2 and 10\,keV determined by a simple power-law fit.}
\tablefoottext{c}{Photon index: $F\propto E^{-\Gamma}$.}
\tablefoottext{d}{Absorption with a equivalent column of hydrogen.}
}
\end{table*}

\subsection{X-ray behavior}

%___________________________________________________________
   \begin{figure*}
   \centering
   \includegraphics[width=15cm]{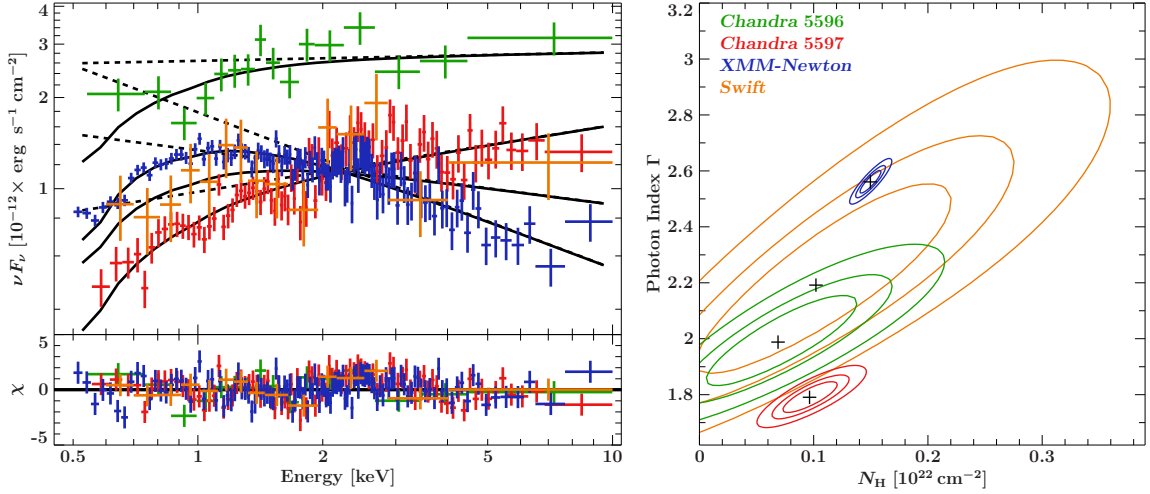}
   \caption{\textit{Left Top panel:} Measured spectral energy distribution of IC\,310 in the X-ray band in different time periods. \textit{Chandra} Obs.\,ID 5596 (green), 
                                     \textit{XMM-Newton} (blue), \textit{Chandra} Obs.\,ID 5597 (red) and \textit{Swift}-XRT (orange). The solid lines show the absorbed
				      power-law fit and the dashed lines show the de-absorbed fit, respectively.
   \textit{Left Bottom panel:} Residuals of the $\chi^2$ fit. 
   \textit{Right panel:} Contour plot between the intrinsic column density $N_{\mathrm{H}}$ and the X-ray photon index $\Gamma$ for the spectra depicted in the left panel. 
                         The contours for each observation are given at 68\%, 90\%, and 99\% confidence level. For comparison, the 
			 Galactic $N_{\mathrm{H}}$ value is $0.12\times10^{22}$\,cm$^{-2}$.}
   \label{XRay1}	
   \end{figure*}
%______________________________________________________________

Table~\ref{table:2} and Fig.~\ref{XRay1} summarize the analysis results of archival X-ray data
for an \textit{XMM-Newton}, two \textit{Chandra} and a \textit{Swift}-XRT observations. The photon
index, $\Gamma$, and the absorption column, $N_\mathrm{H}$, are derived
from fits of power-law models to the 0.5--10\,keV data. Because
intrinsic absorption due to material close to the X-ray source in excess
of the neutral Galactic absorption towards IC\,310 ($N_\mathrm{H}=0.12\times
10^{22}\,\mathrm{cm}^{-2}$; Kalberla et al., \cite{kalberla10}) cannot be excluded,
$N_\mathrm{H}$ is left as a free parameter in these fits.
Note that the obtained $N_\mathrm{H}$ values are compatible with the
Galactic absorption except for \textit{XMM-Newton}.
We also tried an analysis of the \textit{XMM-Newton} data with a broken
power law and a fixed $N_\mathrm{H}$.
It resulted in a good $\chi^2$/d.o.f. but the break energy was found to
be at energies where the absorption takes place, i.e, at $<1$\,keV.
Thus, a break in the spectrum is potentially mimicked by the absorption.

Our analysis shows that between 2003 and 2007,
flux variability is present in the low energy regime (0.5--2 keV) as well as in the higher energy range
(2--10 keV) on timescales of years. This variability is accompanied by changes in the
photon index $\Gamma$ and $N_\mathrm{H}$. A problem with this kind of analysis is that there is a well known correlation between the
parameters, i.e., softer spectra with larger absorption columns cannot
be distinguished from harder spectra with slightly smaller absorption. The
right hand panel of Fig.\ref{XRay1} shows confidence contours between $\Gamma$
and $N_\mathrm{H}$ for all observations. While the \textit{Swift} data have too
low a signal-to-noise ratio to make a statement on changes in
the spectral parameters, a statistically very significant change in
photon index and absorption column is present when comparing the higher
quality \textit{XMM-Newton} and \textit{Chandra} data. The change in the spectral shape cannot
be explained solely by a change in absorption, but the intrinsic
spectrum of the source appears to have changed between the different observations.

We conclude that the intrinsic source spectral index varies varies between $\Gamma=\,2.5$ (soft) and $\Gamma=\,1.8$ (hard). During the
\textit{Chandra} observation the absorption column was consistent with the
Galactic value towards IC\,310, but $N_\mathrm{H}$ was
significantly higher during the \textit{XMM-Newton} observation, which could
be due to an increase in the source intrinsic absorption. A possible cause for a change in the internal absorption of the source could be, e.g., the presence of material 
close to the black hole. Such variation has been seen in a few other active galaxies. For example, a long-term variation
of the intrinsic absorption by 30\% has been observed in Centaurus A
(\cite{benlloch01}), where recent observations also found evidence
for a short term absorption event that lasted for $\sim$170\,d (\cite{rivers12}).

\section{Discussion}

\subsection{Size and energetics of the emission region}

The MAGIC observations establish IC\,310 as a source of variable
VHE $\gamma$-ray emission down to time scales of the order of one day.
The observations rule out the previously discussed origin of the VHE emission at a bow shock driven into the ICM by the radio jets (\cite{neronov}), as this should
produce almost steady-state emission. Causality
allows us to constrain the radius of a spherical emission region from the observed variability time scale $\tau_{\mathrm{var}}$ (in units of days)
\begin{equation}
R\leq 2.6\times 10^{15}\, \tau_{\mathrm{var}}\, \delta\, (1+z)^{-1}\,{\mathrm{cm}},
\end{equation}
where $z$ denotes the cosmological redshift and
$\delta$ the Doppler factor accounting for relativistic bulk motion.
Using the upper limit for the doubling time of the 2009 November flare
with $\tau_{\mathrm{Nov, UL}}=0.55$ and $z=0.0189$, it follows that $R\lesssim\,1.4\times\,10^{15}\,\delta\mathrm{\,cm}$.  
For a black hole mass of  $M_{\mathrm{BH}}\,\simeq 2\times\,10^{8}$\,M$_{\odot}$ inferred from the
central velocity dispersion of IC\,310 \cite{mcelroy1995} and the M-$\sigma$ relation \cite{gultekin2009}, we can compare the size with the Schwarschild radius $R_{\mathrm S}$
yielding 
$R\la 20\,\delta\, R_{\mathrm{S}}$.
Adopting a Doppler factor of $\delta\sim 3-4$ consistent with the properties of the pc-scale radio jet (Kadler et al. \cite{kadler}),
the inferred source size becomes $R\la (60-80)\,R_{\mathrm S}$. This is the scale on which jet formation is expected to take place,
close to the light cylinder of a rotating magnetosphere surrounding the central 
compact object (\cite{mckinney}, \cite{punsly01}). Alternatively, substructures within the jet such as a current-carrying surface sheet (\cite{appl}) or
mini-jets due to magnetic reconnection events (\cite{giannios12}) also could be possible. The emission from small (compared to the
jet diameter) substructures would, however, imply extremely high local photon densities and thus risk
$\gamma$-ray absorption by pair production on the low-energy photon fields.  This would be in conflict with the 
optically thin $\gamma$-ray spectrum up to energies of $\sim$10\,TeV energies.  In turn, the optically thin $\gamma$-ray spectrum
can be used to constrain the local near-infrared (NIR) photon density within the emission region, presumably due to synchrotron radiation of accelerated electrons.
Approximating the $\gamma$-ray energy
$\epsilon_\gamma\simeq 4\, \delta^2 (m_{\mathrm e}c^2)^2/\epsilon_{\mathrm{NIR}}$
where the effective cross section $\sigma_{\gamma\gamma}\simeq 0.2 \sigma_{\mathrm T}$ is maximized for target photons of energy $\epsilon_{\mathrm{NIR}}$,
the condition $\tau_{\gamma\gamma}<1$ leads to the constraint
\begin{equation}
L_{\mathrm{syn}} \la 10^{39}\, \delta^6 \, \tau_{\mathrm{var}}\, {\mathrm{erg}\,\mathrm{s}^{-1} < L_{\mathrm{2\,\mu m}}}=2\times 10^{44}\, \mathrm{erg}\,\mathrm{s}^{-1}
\end{equation}
where $L_{\mathrm{2\, \mu m}}$ denotes the observed NIR luminosity of the
host galaxy \cite{skrutskie}.
Adopting $\delta = 4$, the allowed nonthermal low-energy emission component may reach up to $L_{\mathrm{syn}}\approx 2\%\, L_{2\mu\mathrm m}$
in the NIR, thus allowing only very weak, but still detectable, brightness variations. Such a non-thermal component extending into the X-ray regime
could explain the observed X-ray luminosity. Even if the source of multi-TeV photons is located in the central galaxy, where the isotropic infared photons from the gas and dust torus
of radius $r_{\mathrm dust}$ provide a target for pair production, the photons can escape freely since
$\tau_{\gamma\gamma}(10\,{\mathrm{TeV}}) \simeq 0.025 (L_{20\,\mu\mathrm{m}}/10^{43}\mathrm{erg}\, \mathrm{s}^{-1}) ( r_{\mathrm{dust}}/1{\mathrm{kpc}})^{-1}<1$
interpolating the mid-IR luminosity from the observed luminosities $L_{\mathrm{12\, \mu m}}<9.8\times 10^{42}\, \mathrm{erg\, s^{-1}}$ and
$L_{\mathrm{100\,\mu m}}=4.7\times 10^{43}\, \mathrm{erg}\, \mathrm{s}^{-1}$ (IRAS, \cite{beichman}). If an accretion disk is present, its thermal photons would not significantly increase the optical depth at multi-TeV energies due to the fall-off of the pair production cross section above threshold.\\
Fermi acceleration of electrons with time scale $\tau_{\mathrm{acc}}\simeq 6 r_{\mathrm{L}}/c$
is possible up to the electron Lorentz factors of $\gamma\approx 10^7$
for any plausible magnetic field strength in the jet,
and the synchrotron emission of the accelerated electrons could be responsible for the observed (non-simultaneous) X-ray spectrum (\cite{tammi}).
The observed range of spectral indices is suggestive of an underlying power-law particle distribution with a variable cooling break.
The comoving frame $\gamma$-ray luminosity of $\sim 10^{40}\,\mathrm{erg}\,\mathrm{s}^{-1}$ inferred from the
shock-in-jet scenario amounts to only a small fraction of the total jet luminosity determined from the large-scale radio jet
$L_{\rm jet}=2\times 10^{42}\,\mathrm{erg}\, \mathrm{s}^{-1}$ (\cite{sijbring}), consistent with the observed unperturbed morphology of the radio jet.

\subsection{Comparison with other VHE-emitting AGN}

IC\,310 shows the spectral energy distribution (SED) and short variability time scale
characteristic of blazars, but with an apparent luminosity
of $10^{42-43}$\,erg\,s$^{-1}$, i.e. two to three orders of magnitude lower than typical TeV blazars (Fig.~\ref{FigSED}).
The very hard TeV spectrum of IC\,310 is reminiscent of extreme high frequency peaked BL Lacs (HBL) similar to 1ES\,1426+428 (\cite{wolter08}),
seemingly extending the blazar sequence claimed by Fossati et al. (\cite{fossati98}) and Ghisellini et al. (\cite{ghisellini98}) 
to very low luminosities.
The low apparent luminosity of the blazar in IC\,310 could be the consequence of a relatively large angle between the line of 
sight and the jet.  For example, an angle of $\theta=8.5^\circ$ and bulk Lorentz factor of $\Gamma_{\mathrm{b}}=15$
corresponding to a Doppler factor of $\delta \sim 5$, would imply
that the same source, if viewed under an angle $\theta\sim \Gamma_{\mathrm{b}}^{-1}$ would look one hundred times more luminous.
Indeed, Ghisellini \& Tavecchio (\cite{ghisellini08}) have predicted a large population of weak blazars
with angle to the line of sight $\theta= 4^{\circ}-7^{\circ}$. 

It must be noted, however,  that the X-ray and VHE data discussed 
here are non-simultaneous, and the SED peak values thus uncertain. Therefore, conclusions have to be drawn with caution especially  with respect to flaring episodes present in both data sets. 
The blazar component in IC\,310 dominates at X-to-$\gamma$-ray energies but is only marginal
at infrared-to-optical energies where the host galaxy is at least two orders of magnitude more luminous.  
Giommi et al. (\cite{giommi12a}) argue that $\sim90\%$ of the moderately beamed blazars in the local Universe 
($z<0.07$) are known as radio-galaxies since their blazar components are sub-luminous compared to the host galaxies.
Moderate beaming corresponding to a viewing angle of $10^\circ\la\theta\la 38^\circ$  can indeed explain the
one-sidedness of the jet in IC\,310 (Kadler et al. \cite{kadler}). If there is no bending of the jet, as suggested by
the close match between the observed orientation angles of the radio jet between pc and kpc scales the lower limit restricts the length of the de-projected jet 
from exceeding the extreme end of the jet length distribution of radio galaxies at 1\,Mpc (Neeser et al. \cite{neeser95}). A jet length in excess of 1\,Mpc would also conflict
with the jet life-time of $\sim 10^8$\,yrs (\cite{sijbring}; \cite{deyoung02}).

On the other hand,
the VHE variability of IC\,310 is not faster than the variability of other TeV radio galaxies, e.g. M87 (\cite{aharonian06}; \cite{abramowski12}) 
or NGC\,1275 (\cite{colin12}), which have relatively large angles between the jet and the line of sight. This may indicate that the
bulk Lorentz factor is lower in these radio galaxies than in blazars and that the shock-in-jet scenario becomes inappropriate to accommodate for
the short variability time scales.
Pulsar-type models for emission from the immediate vicinity of the central object (\cite{levinson})
might be a viable option, since anisotropies in the particle and photon distributions
would strongly affect the conclusions drawn in the previous paragraph. However, the observed constancy of the spectral shape is unexpected in this scenario.
Although variations in the accretion rate of an advection-dominated accretion flow and corresponding variations in the gap size can readily explain the flickering behavior
of the $\gamma$-ray flux, the sensitive dependence of the pair-creation optical depth on the accretion rate, and also the energy dependence of the photospheric radius should
lead to notable spectral variability.

Another alternative explanation is $\gamma$-ray emission associated with
cloud passage through the jet, and vice versa, as originally proposed by Blandford \& K\"onigl (\cite{blandford}) to explain variability and
mass entrainment.  Such a scenario has been considered to explain the day-scale VHE flare of M87 as the consequence of proton-proton interactions
in the colliding clouds (\cite{barkov10, barkov12}). Bednarek \& Protheroe (\cite{bednarek97}) proposed stellar-wind-jet interactions
leading to highly anisotropic secondary radiation to explain the rapid $\gamma$-ray variability in AGN jets, and this could also be
of relevance in IC~310 if star formation in its central region can supply the short-lived massive stars with strong winds at a sufficiently high rate (\cite{araudo}).

\section{Summary and Conclusions}

In this paper we presented the re-analysis of the IC\,310 data taken with the MAGIC telescopes between 2009 October and 2010 February.
Using an improved analysis taking into account the data from both wobble positions, we revealed day-scale flux variability for this object.
The size of the emission region is thereby constrained to be smaller than 60-80 Schwarzschild radii of the central supermassive black hole powering
the jet  in IC\,310 (for a black hole mass of $2\times\,10^{8}$\,M$_{\odot}$).

A high and a low state spectrum of the source were defined and investigated independently. The photon indices of both were comparable within the error bars
whereas the flux at 1\,TeV is $\sim$7 times higher in the high state compared to the low state.
The analysis of \textit{Fermi}-LAT data above 10\,GeV from 2008 August to 2011 August shows very faint emission. 
At X-ray photon energies, archival data from \textit{Chandra},
\textit{XMM-Newton} and \textit{Swift}-XRT show clear evidence for flux and spectral variability on times scales of years.
In the X-ray band, both hard and steep spectra with slopes $\Gamma=1.76$--$2.55$ were found, but due to their non-simultaneity, 
it is unclear whether the hard spectral slope is indeed representative of the SED during the $\gamma$-ray observations, as suggested
by comparison with extreme blazar SEDs.  
At $\gamma$-ray energies, the slope of the spectrum measured by \textit{Fermi}-LAT above 10~GeV is $1.3\pm0.5$, while the one
measured by MAGIC above 120\,GeV and reaching almost 10\,TeV is $1.9\pm0.1$.

The day-scale VHE variability rules out emission models occurring in a bow shock between the jet and the ICM, but strongly support the blazar-like scenario.
The SED of IC\,310 can be interpreted as an extreme HBL with synchrotron radiation peaking in the X-ray band
and inverse Compton radiation peaking in the multi-TeV band, although detailed spectral modeling has to await simultaneous data.  
IC\,310 may be considered a representative of a transition population between low-luminosity blazars and FR~I radio galaxies at the faint end of their 
luminosity distributions.

\begin{acknowledgements}
We would like to thank the Instituto de Astrof\'{\i}sica de
Canarias for the excellent working conditions at the
Observatorio del Roque de los Muchachos in La Palma.
The support of the German BMBF and MPG, the Italian INFN, 
the Swiss National Fund SNF, and the Spanish MICINN is 
gratefully acknowledged. This work was also supported by the CPAN CSD2007-00042 and MultiDark
CSD2009-00064 projects of the Spanish Consolider-Ingenio 2010
programme, by grant 127740 of 
the Academy of Finland,
by the DFG Cluster of Excellence ``Origin and Structure of the 
Universe'', by the DFG Collaborative Research Centers SFB823/C4 and SFB876/C3,
and by the Polish MNiSzW grant 745/N-HESS-MAGIC/2010/0.

This research has made use of the NASA/IPAC Extragalactic Database (NED) which is operated by the Jet Propulsion Laboratory, 
California Institute of Technology, under contract with the National Aeronautics and Space Administration.

This research has made use of data obtained from the Chandra Data Archive and the Chandra Source Catalog, and software 
provided by the Chandra X-ray Center (CXC) in the application packages CIAO, ChIPS, and Sherpa. 

Based on observations obtained with XMM-Newton, an ESA science mission with instruments and contributions directly funded by
ESA Member States and NASA.

This research has made use of data obtained from the High Energy Astrophysics Science Archive Research Center (HEASARC), provided by NASA's Goddard Space Flight Center.

The Fermi LAT Collaboration acknowledges support from a number of agencies and institutes for both development and the operation of
the LAT as well as scientific data analysis. These include NASA and DOE in the United States, CEA/Irfu and IN2P3/CNRS in France, ASI and
INFN in Italy, MEXT, KEK, and JAXA in Japan, and the K. A. Wallenberg Foundation, the Swedish Research Council and the National Space
Board in Sweden. Additional support from INAF in Italy and CNES in France for science analysis during the operations phase is also
gratefully acknowledged.

T. Dauser would like to thank the bayerisches Elitenetzwerk and the Bundesministerium
f\"ur Forschung und Technologie under Deutsches Zentrum f\"ur Luft- und
Raumfahrt grant 50OR1207".

We would like to thank the referee for helpful comments.

\end{acknowledgements}

  \institute{ IFAE, Edifici Cn., Campus UAB, E-08193 Bellaterra, Spain
 \and INAF National Institute for Astrophysics, I-00136 Rome, Italy
 \and Universit\`a  di Siena, and INFN Pisa, I-53100 Siena, Italy
 \and Croatian MAGIC Consortium, Rudjer Boskovic Institute, University of Rijeka and University of Split, HR-10000 Zagreb, Croatia
 \and Max-Planck-Institut f\"ur Physik, D-80805 M\"unchen, Germany
 \and Universidad Complutense, E-28040 Madrid, Spain
 \and Inst. de Astrof\'{\i}sica de Canarias, E-38200 La Laguna, Tenerife, Spain
 \and University of \L\'od\'z, PL-90236 Lodz, Poland
 \and Depto. de Astrof\'{\i}sica, Universidad de La Laguna, E-38206 La Laguna, Spain
 \and Deutsches Elektronen-Synchrotron (DESY), D-15738 Zeuthen, Germany
 \and ETH Zurich, CH-8093 Zurich, Switzerland
 \and Universit\"at W\"urzburg, D-97074 W\"urzburg, Germany
 \and Centro de Investigaciones Energ\'eticas, Medioambientales y Tecnol\'ogicas, E-28040 Madrid, Spain
 \and Universit\`a di Udine, and INFN Trieste, I-33100 Udine, Italy
 \and Technische Universit\"at Dortmund, D-44221 Dortmund, Germany
 \and Inst. de Astrof\'{\i}sica de Andaluc\'{\i}a (CSIC), E-18080 Granada, Spain
 \and Universit\`a di Padova and INFN, I-35131 Padova, Italy
 \and Unitat de F\'{\i}sica de les Radiacions, Departament de F\'{\i}sica, and CERES-IEEC, Universitat Aut\`onoma de Barcelona, E-08193 Bellaterra, Spain
 \and Universit\`a  dell'Insubria, Como, I-22100 Como, Italy
 \and Institut de Ci\`encies de l'Espai (IEEC-CSIC), E-08193 Bellaterra, Spain
 \and Tuorla Observatory, University of Turku, FI-21500 Piikki\"o, Finland
 \and Japanese MAGIC Consortium, Division of Physics and Astronomy, Kyoto University, Japan
 \and Inst. for Nucl. Research and Nucl. Energy, BG-1784 Sofia, Bulgaria
 \and Universitat de Barcelona (ICC/IEEC), E-08028 Barcelona, Spain
 \and INAF/Osservatorio Astronomico and INFN, I-34143 Trieste, Italy
 \and Universit\`a  di Pisa, and INFN Pisa, I-56126 Pisa, Italy
 \and ICREA, E-08010 Barcelona, Spain
 \and now at Ecole polytechnique f\'ed\'erale de Lausanne (EPFL), Lausanne, Switzerland
 \and supported by INFN Padova
 \and now at Department of Physics \& Astronomy, UC Riverside, CA 92521, USA
 \and now at Finnish Centre for Astronomy with ESO (FINCA), University of Turku, Finland
 \and also at Instituto de Fisica Teorica, UAM/CSIC, E-28049 Madrid, Spain
 \and now at GRAPPA Institute, University of Amsterdam, 1098XH Amsterdam, Netherlands
 \and Dr-Karl-Remeis-Observatory and Erlangen Centre for Astroparticle Physics, 96049 Bamberg, Germany
 \and Fred Lawrence Whipple Observatory, Harvard-Smithsonian Center for Astrophysics, Amado, AZ 85645, USA
 }

\end{document}